# Significance tests and sample homogeneity loophole

**Authors:** Marian Kupczynski

**Affiliation:**

Département de l'Informatique, Université du Québec en Outaouais (UQO), Case postale 1250, succursale Hull, Gatineau. Quebec, J8X 3X 7 , Canada

Correspondence to: marian.kupczynski@uqo.ca

**Abstract**: In their recent comment, published in Nature, Jeffrey T.Leek and Roger D.Peng discuss how P-values are widely abused in 'null hypothesis significance testing' . We agree completely with them and in this short comment we discuss the importance of sample homogeneity tests. No matter with how much scrutiny data are gathered if homogeneity tests are not performed the significance tests suffer from *sample homogeneity loophole* and the results may not be trusted. For example *sample homogeneity loophole* was not closed in the excellent experiment of Marissa Giustina et al. who reported a significant violation of Eberhard inequality. We are not surprised that Bell type inequalities are violated since if the contextual character of quantum observables is properly taken into account these inequalities cannot be proven. However in order to trust the significance of the violation *sample homogeneity loophole* must be closed. Therefore we repeat after Jeffrey T.Leek and Roger D.Peng that sample homogeneity loophole is probably just the tip of the iceberg.

Sample inhomogeneity may make statistical inference meaningless. In their recent comment[1] Jeffrey T.Leek and Roger D.Peng discuss how P-values are widely abused in 'null hypothesis significance testing' . They underline the importance of data cleaning before exploratory data analysis may start. However, no matter with how much scrutiny data is gathered and cleaned, statistical inference may suffer from *sample homogeneity loophole* (SHL). We explain what SHL is and why it is still not closed in the experiment of Marissa Giustina et al.[2] in which using photons and high-efficiency superconducting detectors they demonstrated the violation of Eberhard inequality without using fair-sampling assumption.

Let us explain SHL on an example of a simple random experiment. A signal produced by some source is sent to a measuring device operating according to unknown internal protocol which produces N successive discrete outcomes forming a finite sample S={ $x_1, x_2, \ldots x_N$ }. In statistics each trial yielding the result $x_i$ is interpreted as a measurement of the value of the corresponding random variable $X_i$ obeying some probability distribution $D_i$. In most general case S is a sample from some time series and has to be studied using specific methods[3,4].

Standard statistical inference and significance tests are reliable only if S is a *simple random sample* drawn from some statistical population what means:

1) all trials and corresponding $X_1,…X_N$ are independent random variables
2) all $X_1,…X_N$ are identically distributed obeying the same probability distribution D.

The violation of the condition 2 can lead to a dramatic breakdown of the statistical inference what we demonstrated in a recent paper[5] written together with Hans De Raedt. If the condition 2 is not carefully checked and sample homogeneity confirmed then the statistical inference suffers from SHL.

In order to show how detrimental SHL can be we quote here one example from our paper[5]. We simulated a random experiment in which a measuring device, operating according to some specific internal protocol, was outputting one of the 6 possible discrete values. We generated 100 runs (each run containing $10^5$ data items). Using these large samples we made a significance test of the null hypothesis $H_0$: 1-B ≥ 0. When three runs 25, 50 and 75 were used the inequality was violated for each run by more than 2000 SEM (standard error of the mean) and one could with great confidence reject the null hypothesis. When we performed the average over 100 runs ($10^7$ data items)   1-B =+0.95 SEM and of course the null hypothesis could not be rejected. The condition 1 was satisfied since all our trials were independent but the samples produced by our device were not homogeneous.

Let us now comment on the experiment of Giustina et al. In this experiment the source is sending polarization-entangled photons which after passing by one of four possible polarization measuring settings $(\alpha_i,\beta_j)$ are sent to two detectors one operated by Alice and another by Bob. The clicks on the detectors are registered and the coincidence counts determined.

Each setting defines a different random experiment and in order to check the Eberhard's inequality one finds the value of a random variable J  being a particular combination of  values of 6 random variables ( 4 coincidence counts and 2 single counts) deduced from the data gathered in all four experimental settings. If the local realism was true J should be always positive thus the null hypothesis tested is $H_0$: J ≥ 0.

The data gathered during 300 seconds of recording per setting were divided into 30 bins and a sample S of 30 different J-values was obtained.  From this sample the value of the mean <J > together with its standard mean error SEM= $s/(n)^{0.5}$ ( n=30, s=sample standard deviation) were estimated and 67σ  (67 SEM) violation of  Eberhard inequality was reported.

Usually we use this terminology if we are convinced that the central limit theorem can be used for our finite sample and  <J >  is normally distributed. Since the normality of this distribution cannot be proven Andrei Khrennikov et al.[7]  used  the Chebychev inequality and concluded that the  null hypothesis can be still rejected at the confidence level 99.95%.

However the null hypothesis test is based on only one (1) sample containing 30 data items and the conditions of a *simple random sample*, were not tested. The violation of independence (condition 1) could simply increase the value of SEM but even if SEM was equal to s the violation of null hypothesis would still be significant.  Since the whole set of data contains the outcomes from 4 different random experiments of course it is not homogeneous  The  sample of observed values of J can only be homogeneous if the data sets obtained for each fixed setting are homogeneous and it was not tested enough.

The experiments testing Bell type inequalities are difficult and several well-known experimental loopholes have to be closed. Giustina et al. closed the " fair-sampling loophole". Johannes Kofler et al.[7] showed that the experiment is immune to the "production-rate loophole" and that the results are consistent with quantum theory. Jan-Ake Larsson et al.[8] proved that the experiment is not vulnerable to the "coincidence-time loophole". It is clear that it is an excellent experiment performed with great scrutiny. Nevertheless unless the additional sample homogeneity tests are performed SHL is not closed and its impact on conclusions of the significance test is unknown.

From the discussion above one could understand that we do not believe in the experimental evidence of the violation of Bell type inequalities. It is just the opposite. In our opinion [9,10], if the contextual character of quantum observables is properly taken into account, the Bell type inequalities cannot simply be proven thus there is no reason to believe that they cannot be violated. The violation of these inequalities does not prove a mysterious quantum nonlocality but it only proves the incorrectness of counterfactual reasoning and probabilistic models used to prove them. This point of view is shared by several authors[11-15]. By no means is the list of the references complete and other references may be found in the cited papers and books.

As we mentioned before we have chosen Giustina et al. experiment only as an example. There are many excellent experiments in different domains of science where SHL was not closed or could not be closed. Since SHL can be detrimental for the significance tests we can repeat after Jeffrey T.Leek and Roger D.Peng that sample homogeneity loophole is probably just the tip of the iceberg.

To be on safe grounds sample homogeneity tests should be incorporated in statistical analysis of any large sample of experimental data.

**References**

1. Leek, J.T and Peng, R.D, Statistics: P values are just the tip of the iceberg, *Nature* **520**, *612 (30 April 2015) doi:10.1038/520612a*
2. Giustina, M. *et al* , Bell violation with entangled photons, free of the fair-sampling assumption, *Nature* **497**, 227-230 (2013)
3. Box, G.E.P, Jenkins, G.M. and Reinsel, G.C., *Time Series Analysis Forecasting and Control*, Hoboken: Wiley (2008)
4. Kupczynski, M., Time series, stochastic processes and completeness of quantum theory, *AIP. Conf. Proc.* **1327**, 394 -400 (2011)
5. Kupczynski, M. and De Raedt, H., Breakdown of statistical inference from some random experiments, arXiv:1307.6475 [quant-ph]
6. Khrennikov, A. Yu.et al, On the equivalence of the Clauser–Horne and Eberhard inequality based tests , *Phys. Scr.* **2014** 014019 doi:10.1088/0031-8949/2014/T163/014019
7. Kofler, J., Ramelow,S., Giustina,M. and Zeilinger,A, On 'Bell violation

using entangled photons without the fair-sampling assumption'. arXiv:1307.6475 [quant-ph]
8. Larsson, J.-A. *et al*, Bell violation with entangled photons, free of the coincidence-time loophole, *Phys. Rev. A* **90**, 032107 (2014)
9. Kupczynski, M., Causality and local determinism versus quantum nonlocality. *J. Phys.: Conf. Ser.* **504** (2014) 012015. doi:10.1088/1742-6596/504/1/012015
10. Kupczynski, M., Bell Inequalities, Experimental Protocols and Contextuality, *Found. Phys*. (12 Dec 2014), doi:10.1007/s10701-014-9863-4
11. Nieuwenhuizen, T. M., Where Bell went wrong, *AIP Conf. Proc*. **1101**, 127-33 (2009)
12. Khrennikov, A. Yu., *Contextual Approach to Quantum Formalism*, Springer, Dortrecht (2009)
13. Khrennikov, A. Yu., Classical probability model for Bell inequality, *J. Phys.: Conf. Ser.* **504** (2014) 012019 doi:10.1088/1742-6596/504/1/012019
14. Hess, K., De Raedt, H. and Michielsen. K., Hidden assumptions in the derivation of the theorem of Bell. *Phys. Scr.* **T151**, 014002 (2012)
15. Żukowski, M and Brukner, Č., Quantum non-locality—it ain't necessarily so. *J. Phys. A: Math. Theor*. **47** (2014) 424009 (10pp). doi:10.1088/1751-8113/47/42/424009